\documentstyle[prl,aps,psfig]{revtex}
\begin{document}
\draft
\preprint{}


\title{Spheres and Prolate and Oblate Ellipsoids from an 
Analytical Solution of Spontaneous Curvature Fluid Membrane Model}

\author{Quan-Hui Liu$^{1,2}$\cite{l}, Zhou Haijun${^1}$, Ji-Xing Liu${^1}$, Ou-Yang Zhong-Can${^1}$}

\address{$^{1}$Institute of Theoretical Physics, Academia Sinica, P.O. Box 2735, Beijing, 100080, China}
\address{$^{2}$ Department of Physics, Hunan University, Changsha, 410082, China}
\date{\today}

\maketitle


\begin{abstract}
An analytic solution for Helfrich spontaneous curvature membrane model
(H. Naito, 
M.Okuda and Ou-Yang Zhong-Can, 
Phys.\ Rev.\ E {\bf 48}, 2304 (1993); {\bf 54}, 2816 (1996)),  which 
has a conspicuous feature of representing
the circular biconcave shape, is studied. Results show that the solution in fact describes a family of shapes, 
which can be classified as: i) the flat plane (trivial case), ii) the sphere, 
iii) the prolate ellipsoid,
iv) the capped cylinder,
v) the oblate ellipsoid,
vi) the circular biconcave shape,
vii) the self-intersecting inverted circular biconcave shape,
and viii) the self-intersecting nodoidlike cylinder.
Among the closed shapes (ii)-(vii), a circular biconcave shape is the one with the minimum of 
local curvature energy.
\end{abstract}


\section{Introduction}
Why red blood cells under normal 
physiological conditions take the circular biconcave shape (CBS) but not the spherical shape 
aroused the long standing curiosity of human being since its first discovery in the seventieth 
century. Today, physicists ascribe it to the minimization of bending energy of flexible 
lipid bilayer membrane consisting of amphiphilic molecules.
The first successful model revealing the morphology was due to Canham\cite{cah} in
1970. However, his theory suffered from the shortcoming that 
the membrane was assumed to consist of 
two identical labile surfaces, and both chemical and physical environments between
two sides of the membrane were assumed to be identical too. To overcome the shortcoming,  
Helfrich in 1973 introduced a phenomenological
parameter, namely the spontaneous curvature, in describing the more realistic situations:
the asymmetry of the two leaflets of the membrane and the chemical or/and physical differences 
between the interior and exterior membrane\cite{hel}. As expected, this model gave more abundant shapes 
than that of Canham. Based on the numerical integration technique\cite{deu1}, Helfrich 
spontaneous curvature model yielded a catalog of
axisymmetric vesicle shapes: the CBS, the prolate and oblate ellipsoid, etc.
More than ten years later,
the general equilibrium equation was derived by performing the variation of the Helfrich energy 
functional\cite{oy89}. The first triumph of the equation was the prediction of the existence 
of Clifford torus shape membrane
\cite{oy91} and its subsequent experimental verification\cite{mut}.  In 1993, the general equation 
in axisymmetric case was obtained, and it is a complicated 
third order nonlinear differential equation \cite{huoy}. An analytic solution capable of representing 
the CBS\cite{oy93} was immediately obtained. In the same year, the first integral of the third order nonlinear 
differential equation was found; then the equation reduced to be a second order one 
involving a constant of integration $C$\cite{zhe}. An interesting fact is that
the analytical solution requires a nonvanishing value of $C$\cite{oy93,zhe}, 
whereas only the special case of the second order differential equation with $C=0$ has been well studied by numerical methods\cite{serf}. 
In 1996, the solution was used to explain the experimentally observed polygonal shape 
transformation of the CBS\cite{oy96}. Studies in this paper will show that this solution actually represents a family of  
shapes in this solution, but in the family, the CBSs have the lower energies and a CBS has the lowest.
For convenience, we will call this solution the CBS solution hereafter.

The existence of the analytical CBS solution in Helfrich model proves to be
remarkable, because the analytical solutions to a nonlinear theory are rare 
and precious\cite{oy95}. 
In very special cases of $c_0=\delta p=\lambda=0$ (cf. following Eq.(1)), 
the Helfrich energy functional reduces to that for
constant curvature surfaces and Willmore surfaces\cite{nic}, and all known analytical solutions 
having physical applications in membrane shapes are Delaunay surfaces, sphere, torus\cite{nic}, and 
no more\cite{pin}. 
Therefore a systematic study of the CBS solution is necessary. We will find that there are eight types of  
shapes involved in the solution and the enclosed shapes
can be grouped into the prolate or oblate ellipsoid branches. These two branches are bifurcated from a sphere.

The article is organized as follows. In section II, how to obtain the CBS solution from the Helfrich model 
is outlined. In section III, all typical shapes contained in the CBS solution
are plotted and their parameterizations are presented. In section IV, a systematically analysis 
of the CBS solution is given. In the shape family, there is an shape with minimum energy and in section V
this shape is found from the the scale invariance of the local curvature energy. 
In section VI, a comparison of our results with the previous 
experimental and theoretical results is given. In the final section VII, a brief conclusion is given.

\section{Helfrich Spontaneous Curvature Model and its CBS Solution}

  Equilibrium shapes of phospholipid vesicles are assumed to correspond to 
the minimum of the elastic energy of the closed bilayer membrane. The energy functional of Helfrich
spontaneous curvature model reads as\cite{hel}
\begin{equation} 
F={\frac{1}{2}} k\int (c_1+c_2-c_0)^2 dA+\delta p\int dV+\lambda \int dA,
\label{e1}
\end{equation}
where $dA$ and $dV$ are the surface area and the volume element for the vesicle, 
respectively, $k$ is an 
elastic modulus, $c_1$ and $c_2$ are the two principal curvatures of the surface, 
and $c_0$ is the spontaneous 
curvature to describe the possible asymmetry of the bilayer membrane. When $c_0$ is zero Helfrich
model reduces to Canham model\cite{cah}. The Lagrange multipliers $\delta p$ and $\lambda$ take account of the 
constraints of constant volume and area, which can be physically understood as the osmotic pressure between 
the ambient and the internal environments, and the surface tension, respectively. 
The general equilibrium shape equation is\cite{oy89}
\begin{equation}
 \delta p-2\lambda H+k(2H+c_0)(2H^2-2K-c_{0}H)+ 2 k{\bigtriangledown}^2 H=0,
\label{e2}
\end{equation}
where ${\bigtriangledown}^2={1 \over \sqrt{g}}{\partial}_{i}(g^{ij}\sqrt{g}{\partial}_{j})$ is 
the Laplace-Beltrami operator, $g$ is the determinant of the 
metric $g_{ij}$ and $g^{ij}=(g_{ij})^{-1}$, $K=c_1 c_2$ is the Gaussian curvature and 
$H=-(1/2)(c_1+c_2)$ is the mean 
curvature. Assuming that the shape has axisymmetry, the general shape equation Eq.(2) becomes a 
third order nonlinear differential equation\cite{huoy}
 \begin{eqnarray}
 \cos^{3}\psi(\frac{d^{3}\psi}{dr^{3}}) & = & 4\sin\psi\cos^{2}\psi(
\frac{d^{2}\psi}{dr^{2}})(\frac{d\psi}{dr}) - \cos\psi(\sin^{2}\psi-
\frac{1}{2}\cos^{2}\psi )(\frac{d\psi}{dr})^{3} \nonumber \\
 & + & \frac{7\sin\psi\cos^{2}\psi}{2r}(\frac{d\psi}{dr})^{2} - 
\frac{2\cos^{3}\psi}{r}(\frac{d^{2}\psi}{dr^{2}}) \nonumber \\
 & + & [\frac{c_{o}^{2}}{2} - \frac{2c_{o}\sin\psi}{r} + \frac{\lambda}{k}
 - \frac{\sin^{2}\psi-2\cos^{2}\psi}{2r^{2}}]\cos\psi(\frac{d\psi}{dr})
 \nonumber \\
 & + & [\frac{\delta  p}{k} + \frac{\lambda\sin\psi}{kr} + 
\frac{c_{o}^{2}\sin\psi}{2r} - \frac{\sin^{3}\psi+2\sin\psi\cos^{2}\psi}
{2r^{3}}],
 \label{oyhu}
\end{eqnarray}
where $r$ is the distance from the symmetric $z$ axis of rotation,
$\psi(r)$ is the angle made by the surface tangent and the $r$ axis as shown in FIG.1.
The positive direction of the angle is that the angle measured clockwise from $r$ axis.
It is contrary
to the usual mathematical convention; therefore, the mean curvature $H$ is $-1/2(\sin\psi/r+d\sin\psi/dr)$, in which
 $ c_1=\sin\psi/r $ denotes the principal curvature along the parallels of latitude, and 
$c_2=d\sin\psi/dr$ denotes that along those of meridian. It is
worthy to mention that the spontaneous curvature $c_0$ carries a sign. When the normal of a surface
change its direction, $c_1, c_2$ and $c_0$ must change their signs simultaneously. 
Keeping the directions of $r$ and $z$ as the usual, we have consequently 
\begin{equation}
\left\{
\begin{array}{l}
dz/dr=-\tan\psi(r) \nonumber \\ 
z(r)-z(0) =-\int ^{r}_0 \tan\psi(r) d r\\
{\bf n}=(\sin\psi \cos\phi , \sin\psi \sin\phi ,\cos\psi )
,
\end{array}
\right. 
\label{normal}
\end{equation} 
where ${\bf n}$ denotes the normal of the surface and $\phi $ the azimuthal angle.


For self-consistence, the positive direction of arclength $s$ along the contour 
in $r-z$ plane is necessary to start from the north pole of the shape.
 If so, we see that the parameterization for sphere is  
$\sin\psi(r)=r/R_0$ where $R_0>0$ is the radius of the sphere, 
and our sign convention used in this paper is then
compatible with that used in most previous works\cite{deu1,huoy,serf}. 
The third order nonlinear differential equation (3)
can be simplified to be a second order one\cite{zhe}
\begin{eqnarray}
&&\cos^2\psi \frac{d^2\psi}{dr^2}
-\frac{\sin(2\psi)}{4}(\frac{d\psi}{dr})^2
+\frac{\cos^2\psi}{r}\frac{d\psi}{dr}-
\frac{\sin(2\psi)}{2 r^2} \nonumber \\ 
&&-\frac{\delta p r}{2 k \cos\psi}
-\frac{\sin\psi}{2 \cos\psi}(\frac{\sin\psi}{r}-c_0)^2
-\frac{\lambda \sin\psi}{k \cos\psi}=\frac{C}{r \cos\psi},
\end{eqnarray}
where $C$ is a constant of integration. It is still a rather complicated equation  
and does not belong to any known type of well-studied differential equation in mathematics. 

   Under the conditions that both the surface tension $\lambda$ and the osmotic pressure difference 
$\delta p$ are zero, i.e., 
\begin{equation}
\lambda=\delta p =0,
\end{equation}
an analytic CBS solution for the Eq.(3) \cite{oy93,oy96}, or Eq. (5) with $C=2 c_0$, is
\begin{equation}
\sin\psi=r/R_0+ c_0 r \ln r,~~{\rm or~equivalently},~~\sin\psi=c_0 r \ln(r/r_m), 
\label{sin}
\end{equation}
where $R_0$ is an arbitrary constant and $r_m=\exp(-1/(c_0R_0))$.
When we first obtained this solution\cite{oy93,oy96}, we concentrated on the fact that it can be used to represent the 
CBS of usual red blood cells\cite{oy96}. In fact, by adjusting parameter $c_0$ in the interval $(-\infty,\infty)$,  
the formula (\ref{sin}) can give a family of shapes, 
which is what we are going to analyze in detail. For sake of convenience, we mainly use the first form of the CBS 
solution (\ref{sin}) with $R_0=1$. 

\section{All Possible Shapes in the CBS Solution }

Before presenting the shapes, we would like to make two comments on how to characterize 
the shapes. First, for each shape, we will give both the parameterization and the interval of $r\in[0,\infty)$ in which the shape appears. 
This is because that there may be different shapes represented by the same parameterization in distinct intervals of 
$r\in[0,\infty)$ as long as in the intervals $|\sin\psi(r)|\leq 1$ are satisfied.  
Second, even the scale invariance will be fully studied in section V, 
we will follow the usual usage\cite{serf} to give for each shape the scale invariant $c_0 r_s$, in which $r_s=\sqrt{A/4 \pi}$ with $A$ denoting the 
area of the surface of the shape.  
Unless specially mentioned and discussed, that the normals of the surface point outwards for closed shapes is always implied. 

All possible types of shapes represented by Eq. (\ref{sin}) are listed in the following. \par
i) The flat plane (trivial case) with $c_0 r_s=0$:
\begin{equation}
\sin{\psi}=0,\qquad r \in[0,\infty].
\end{equation}

ii) The sphere with $c_0 r_s=0$:
\begin{equation}
\sin{\psi}=r/R_0,\;\;r \in[0,R_0].
\end{equation}

iii) The prolate ellipsoid. A typical shape with $c_0 r_s=-0.72$ is shown in FIG.2(a) 
and one of its parameterization is
\begin{equation}
\sin{\psi}=r-0.6 r\ln r,\;\;r\in[0,1].
\end{equation}
Another parameterization representing the same shape is $\sin{\psi}=r-1.85 r\ln r$, $r\in[0,0.324]$. 
The parameterization $\sin{\psi}=r+4.063 r\ln r$, $r\in[0,0.148]$ represents the same shape but with inwards pointing normal.

iv) The capped cylinder.  A typical shape with $c_0 r_s=-2.06$ is shown in FIG.2(b) and one of its parameterization is
\begin{equation}
\sin{\psi}=r-0.99r\ln r,\;\;r\in[0,1].
\end{equation}
Another parameterization representing the same shape is $\sin{\psi}=r-1.01 r\ln r$, $r\in[0,0.980]$. 
The parameterization $\sin{\psi}=r+3.5913 r\ln r$, $r\in[0,0.275]$ represents the same shape but with inwards pointing normal.

v) The oblate ellipsoid.  A typical shape with $c_0 r_s=0.46$ is shown in FIG.2(c) and its
parameterization is
\begin{equation}
\sin{\psi}=r+0.5 r\ln r,\;\;r\in[0,1].
\end{equation}

vi) The CBS. A typical shape with $c_0 r_s=1.51$ is shown in FIG.2(d) and its parameterization is
\begin{equation}
\sin{\psi}=r+1.8 r\ln r,\;\;r\in[0,1].
\end{equation}

vii) The self-intersecting inverted CBS. A typical shape with $c_0 r_s=2.72$ 
is shown in FIG.2(e) and its parameterization is 
\begin{equation}
\sin{\psi}=r+3.2 r\ln r,\;\;r\in[0,1].
\label{selcb}
\end{equation}
To note that in this case, the outside of this shape can be defined but we can not let the normal always point outwards 
using a single parameterization (\ref{selcb}). 
The normal of the parameterization (\ref{selcb}) points outwards only at the surface of the torus part.  

viii) The self-intersecting nodoidlike cylinder. This situation was discussed
in previous paper\cite{oy95}. We add a typical figure, FIG.2(f), in this paper for 
completeness. The cylinder is infinitely long with periodic packing, along the rotational axis, of a basic unit in which self-intersecting occurs once. 
We plot the basic unit only. We use the surface of the basic unit to calculate $r_s$ and the shape has $c_0 r_s =3.28$. 
Its parameterization is
\begin{equation}
\sin{\psi}=r+3.60r\ln r, \;\;r\in[0.301,1].
\end{equation}
The normal of this parameterization points outwards only at the surface of the torus part also. To 
note that the same parameterization $ \sin{\psi}=r+3.60r\ln r\ $ in interval 
$ r\in [0,0.257] $ gives a capped cylinder but its normal points inwards. 


These eight shapes consist of all possible types of
shapes containing in the solution. From the parameterizations of the these shapes, one
may have noticed two facts that the spontaneous curvatures of all shapes are within a very narrow domain instead of 
infinite domain of $c_0\in (-\infty, \infty)$,  and a single shape may have different parameterizations. 
Explanations of these facts will be given in next section by both qualitative and quantitative studies 
of the CBS solution. 

\section{A quantitative study of shapes in the CBS solution}

All the closed shapes (ii)-(vii) presented in last section can be grouped into two branches: 
The prolate ellipsoid including (ii)-(iv) and oblate ellipsoid branch including (ii), (v)-(vii), and
these two branches are bifurcated from a sphere (ii). It is easily understandable
if looking into the CBS solution (\ref{sin}) $\sin\psi=r/R_0+ c_0 r \ln r$.
When $R_0>0$ and $c_0=0$, this 
form gives nothing but sphere of radius $R_0$. Then when $c_0$ is a small quantity, positive
$c_0$ leads to oblate ellipsoid and negative $c_0$ leads to prolate ellipsoid respectively. 
What if $c_0$ is large is not so evident and needs some reasoning. 

For analyzing how $c_0$ affects the shapes and the intervals of $r$ such that $|\sin\psi (r|\leq1$, we use the parameterization  
$\sin{\psi}=r+c_0 r\ln r$, and sketch the relation of $\sin{\psi}$ with positive $c_0$ vs. $r$ in FIG.3. 


From this figure, we see that $\sin{\psi}=r+ c_0r\ln r$ is a single-valued and monotonous function of $r$; and it reaches its 
extremum at point $r_2$ which satisfies
\begin{equation}
\frac{d\sin{\psi}}{dr}=1+c_0+ c_0\ln r_2=0,\; 
i.e. ,\;\;r_2=\exp(-\frac{1+c_0}{c_0}).
\end{equation}
The extremum value of $\sin\psi(r_2)$ is
\begin{equation}
\sin\psi(r_2)=-r_2 c_0=-c_0 \exp(-\frac{1+c_0}{c_0}),
\label{max}
\end{equation}
which is negative for positive $c_0$ and vice verse. Taking the positive  
$c_0$ as an example in FIG. 3, corresponding to different $c_0$, shapes can appear in different intervals of $r$. 
There are three different cases of $c_0\in [0, \infty)$.  
i) When $c_0$ is critical such that $\sin\psi(r_2)= -1$, we have $c_0=c_r\approx 3.59112$ from Eq.(\ref{max}). 
The meridian principal curvature vanishes at point $r_2$ as 
$c_2=d\sin{\psi}/{dr}=0$: it is the unphysically infinitely long capped cylinder with radius $r_2=1/c_0=0.278$ from Eq. (\ref{max}).
ii) When $c_0<c_r$, we have $\sin\psi(r_2)>-1$ and shapes appear in one interval 
$[0,r_4]$ in which $r_4$ are the equatorial radii of the shapes such that $\sin\psi(r_4)=1$. 
Along the increase of $r$ from $0$ to $r_4$, the tangent angle $\psi(r)$ decreases from zero and reaches its minimum at 
point $r_2$, then increases monotonously and reaches its maximum value $\psi=\pi /2$ at point $r_4$.  
These shapes certainly belong to the oblate ellipsoid branch, and their normals point outwards from Eq. (\ref{normal}). 
iii) When $c_0>c_r$, we have $\sin\psi(r_2)<-1$, shapes can appear in two
distinct intervals $[0,r_1]$, in which $r_1$ are the equatorial radii of the shapes such that $\sin\psi(r_1)=-1$,  
and $[r_3, r_4]$, because in both intervals $|\sin\psi(r)|\leq 1$ is satisfied. In interval $[0,r_1]$, 
the tangent angle $\psi(r)$ decreases from zero and terminates at $\psi (r_1) =- \pi/2$. These shapes certainly belong to 
the prolate ellipsoid branch, but the normals point inwards since $-1\leq \sin\psi(r)\leq 0$ holds for whole interval $r\in [0,r_1]$.  
In distinct interval $[r_3, r_4]$, numerical studies show that only the self-intersection nodoidlike cylinder appears. 
 
In fact, using the parameterization 
$\sin\psi=r/R_0+ c_0 r \ln r$, each shape with a $c_0$ outside the domain $(-1,c_r)$ 
can be found an identical shape with a value of $ c_0 $ within the domain $ (-1, c_r) $. For demonstrating this fact, 
we resort to the second form of the CBS solution $ \sin \psi=c_0 r \ln (r/r_m) $ in Eq. (\ref{sin}).  Two parameterizations 
$c_0 r \ln (r/r_m)$ and $- c_0 r \ln (r/r_m)$ represent the same shape, but one's normal is
opposite to another's from Eq.(\ref{normal}). If requiring that the normal points outwards, one of these two parameterizations must be
eliminated. Furthermore we have a simplified form 
\begin{eqnarray}
\sin\psi &=&c_0 r \ln(r/r_m) \nonumber\\
         &=&(c_0 r_m)(r/r_m)ln(r/r_m)\nonumber\\
         &=&c'_0 xlnx,
\label{dim}
\end{eqnarray}
where $x=r/r_m$ is the dimensionless length, and $ c'_0=c_0 r_m=c_0 exp(-1/(c_0R_0)) $ is a dimensionless spontaneous curvature.
One can find that the $c'_0$ in two separate domains $c'_0\in (-\infty, -e)$ 
and $c'_0\in [0,e)$, in which $e$ is the base of the natural logarithm, suffice to give all possible shapes with normals pointing 
outwards in the CBS solution. When $c'_0=\pm e$, the shapes are unphysically infinitely long capped cylinders. 
From the relation $ c'_0=c_0 exp(-1/(c_0R_0)) $ 
and letting $R_0=1$, all shapes can be mapped into a single domain $c_0\in (-1, c_r)$ in parameterization 
$\sin\psi=r/R_0+ c_0 r \ln r$. This is why there are only a few of spontaneous curvature $c_0$ within a narrow domain in the last section 
are sufficient to give all possible types of shapes in the CBS solution, i.e., why each shape with a $c_0$ outside the domain $(-1,c_r)$ 
can be found an identical shape with a value of $ c_0 $ within the domain $(-1, c_r)$.

In addition to the above qualitative analyses, numerical method will be used to characterize quantitatively the shapes 
in whole domain $c_0 \in (-\infty,\infty)$. 
We introduce two ratios, semi-axis ratio $z(r_0)/r_0$ and reduced radius ratio $r_v/r_s$. The so-called  semi-axis
$z(r_0)$ is defined by the half of the distance of two poles of a vesicle in the symmetrical axis $z$, 
which is
\begin{equation}
z(r_0)-z(0) =-\int ^{r_0}_0 \tan\psi(r) d r,
\end{equation}
where $r_0$ is the equatorial radius of the vesicle satisfying $\sin\psi(r_0)=\pm1$. The ratios $z(r_0)/r_0$ are plotted by
dashed lines in FIG. 4. From FIG. 4(b), the $c_0$ approaches $-1$ and $c_r$, the ratio $z(r_0)/r_0$ approaches
infinity: the shapes are infinitely long capped cylinders with radii $1$ and $0.278$ respectively. When $c_0$
increases from $-1$ to $c_r$, the ratio decreases monotonously. We have in sequence the capped cylinder, 
the prolate ellipsoid and the sphere of unit radius at $c_0=0$, the oblate ellipsoid with a small positive $c_0$, 
the CBS and the center of the CBS touches at $c_0=2.4288$,
then self-intersecting inverted CBS.  From FIG. 4(a) and
FIG. 4(c), we clearly see that when $c_0<-1$ or $c_0>c_r$, the ratios $z(r_0)/r_0$ are greater than $1$. 
We have in sequence the capped cylinder, the prolate ellipsoid and the quasi-sphere with increasing $|c_0|$. 
When $|c_0|$ approaches infinity, the constraint $|\sin\psi(r)|\leq 1$ means that $r_1$, cf. FIG. 3,  must be very small. 
Thus we have approximately $\sin\psi(r)\simeq \mp c_0 r$ for positive and 
negative $c_0$ respectively using L' Hospital method. It implys that we will have sphere again. 
All these shapes are closed and appear in intervals of $r$ including the point $r=0$, which are physically interesting and 
are plotted in the FIG. 4.
When $c_0<-1$ or $c_0>c_r$, in intervals $[r_3>0,r_4]$ (cf. FIG. 3), we have the 
self-intersecting nodoidlike cylinders, which are less physically interesting and have not been plotted in the FIG. 4.
As pointed out in \cite{oy95}, along with the increasing $|c_0|$, the number of self-intersections 
in unit length increases and it approaches a quasi-torus as $|c_0|$ approaches $\infty$. In fact,
when $-1<c_0\leq 0$, the self-intersecting nodoidlike also appear in the intervals $[r_3>0,r_4]$, but
along with the decreasing $|c_0|$, the number of self-intersections in unit length increases.

The second ratio is $r_v/r_s$, in which $r_v$ is defined by the radius of the sphere having the same volume $V$
of a vesicle and $r_s$ the radius of the sphere having the same area $A$ of the vesicle. The volume $V$ and 
area $A$ are
\begin{eqnarray}
V&&=4\pi \int ^{r_0}_{0}\frac{r(z(r)-z(0))}{\cos\psi} d r,\\
A&&=4\pi \int ^{r_0}_{0}\frac{r}{\cos\psi} d r.
\end{eqnarray}
Undoubtedly, we have $r_v/r_s\leq 1$ and the equality 
holds for sphere only. The ratio is plotted by dot-dashed lines in FIG.4. 

The scale invariant $c_0 r_s$ is plotted by thin solid lines in FIG. 4. In FIG. 4(c), we plot both $c_0 r_s$ 
(above the $c_0$ axis) and $-c_0 r_s$ (below the $c_0$ axis ) by the same thin solid lines, in order to see
the fact that each shape with a $c_0\in (-\infty,-1)$ is identical to a shape with a $c_0\in (c_r, \infty)$ and vice verse.
But the normals of shapes parameterized by $\sin\psi=r+c_0 r \ln r$ with $c_0\in(c_r,\infty)$ point inwards.  

From both the qualitative and quantitative studies in this section, we can draw a conclusion that all possible closed shapes in 
the CBS solution (8) are, along with $c_0$ increasing from $-1$ to $c_r$, the capped cylinder, the prolate ellipsoid, the sphere, 
the oblate ellipsoid, 
the CBS, and the self-intersecting inverted CBS.


\section{The scale invariance of the energy and the shape with minimum energy }

An important property of the local curvature energy  is its
scale invariance. This energy does not depend on the size of the vesicle but only on its shape\cite{serf}. 
If $\bf{R}$ is a solution with a local curvature energy, the rescaled shape ${\bf R}\rightarrow {\bf R} /k$ with $k>0$, 
and consequently $(c_1, c_2, c_0, dA)\rightarrow (k c_1, k c_2, k c_0, dA/k^2)$, is also a solution with the 
same local curvature energy.
Two parameterizations of the CBS solution (8) having manifest scale invariance are
\begin{equation}
\sin\psi(r)=r/R_0+c_0 r \ln (r/r_0), \;\; and \qquad \sin\psi(r)=c_0 r \ln (r/r_0),
\end{equation}
in which $r_0$ is a constant with length unit as $r$ or $R_0$. It is evident that the two products $c_0 r_s$ and $c_0 r_v$ 
are two scale invariants independent of a particularly chosen parameterization. 
We used two particular parameterizations, $\sin\psi(r)=r/R_0+c_0 r \ln r$ and  
$\sin\psi(r)=c_0 r \ln r$, to search for the minimum energy shape. Both gives the same result $c_0r_s=1.04$ and $c_0 r_v=1.00$.
The minimum energy is $0.480$; hereafter we take the energy of sphere $8 \pi k$ as the energy unit. 
Surprisingly, the CBS with touching center at $c_0=2.4288$ has energy $1.00$. 
In FIG. 4 (b), the thick solid line shows that there is a energy minimum shape with $c_0=1.200$: it is 
a CBS. In general, the oblate ellipsoids including the CBS have lower 
energy than other shapes including the sphere, prolate ellipsoid and the self-intersecting inverted CBS have.
 
\section{A comparison of our results with the previous experimental and theoretical results}
  Even all the previous numerical studies concentrated on the solution of equation when 
$C=0$ of Eq. (5), the obtained shapes \cite{serf,dob} included all the eight shapes above.
And also, all these eight shapes except the self-intersecting cases have been observed 
in laboratory\cite{dob,hot,bess}. 

  Since our solution belongs to the situation $C\not=0$ of Eq.(5), our results coincide
with the known theoretical results in one respect but differ from in the other. 
First, our approach supports the conclusion that the prolate ellipsoids 
may have higher energy than the oblate ones have\cite{oy89}. However,
the standard instability analysis of a sphere starts from a slightly deformed sphere of 
parameterized form $r=R_0+\sum a_{lm}Y_{lm}(\theta,\phi)$ where $a_{lm}$ is a set of small 
parameters corresponding to spherical harmonics $Y_{lm}(\theta,\phi)$. For small $a_{lm}$, for example the only nonvanishing $a_{2,0}$ satisfying 
$|a_{2,0}|<<R_0/2$, the ellipsoids have 
$C^{\infty}$ contour curves. But our ellipsoids have only $C^{1}$ contour curves, 
We must stress that the $C^{1}$ continuity suffices to ensure that the membranes are free from any force 
acting on the any point, otherwise the 
contour curve may be $C^{0}$. The 
second derivative of $z(r)$ with respect to $r$ is singular at point $r=0$. One may feel uneasy of this singularity. 
In fact, it relates to energy density only, and the total energies $(1/2) k\int (c_1+c_2-c_0)^2 dA$ are limited
as shown in FIG.4. Second, the infinitely long capped cylinders appear
when $c_0=-1$ and $c_0=c_r$, and they correspond the infinite energy states, as shown by thick
solid lines in FIG.4 (a)-(c). From our results, the phases between the two sides 
of $c_0=-1$ and $c_0=c_r$ are separate. The latter $c_r$ distinguishes
a oblate ellipsoid and a prolate ellipsoid phase; it is the same situation in the
known phase diagram\cite{serf}. However when $c_0$ 
changes from $c_0<c_r$ to $c_0>c_r$, the normal changes its direction from pointing
outwards to pointing inwards. It seems that the membrane takes a global flip-flop
procedure. But no such procedure takes place when $c_0$ 
changes from $c_0<-1$ to $c_0>-1$ even the transitions are also discontinuous.
 Third, there is an interesting number appeared both 
in our results and the previous ones, $c_0 R_0=1.200$. We see that from the FIG. 4(b)
when $c_0R_0=1.200$ where $R_0=1$ in our parameterization, the CBS
has the minimum energy. In the usual instability analysis of a sphere via infinitesimal deformation, 
the infinitesimally deformed oblate shape
has the lower energy and more stable than the infinitesimally deformed prolate one whenever $c_0R_0<-1.2$\cite{oy89}.

\section{Conclusion}

An analytic solution for Helfrich spontaneous curvature membrane model
\cite{oy93,oy96},  which 
has a conspicuous feature of representing
the circular biconcave shape, is systematically studied in this paper. Results show that the solution in fact describes a family of shapes, 
 which can be classified as: 
i) the flat plane (trivial case),
ii) the sphere,
iii) the prolate ellipsoid (FIG.2(a)),
iv) the capped cylinder (FIG.2(b)),
v) the oblate ellipsoid (FIG.2(c)),
vi) the CBS (FIG.2(d)),
vii) the self-intersecting inverted CBS (FIG.2(e)) and 
viii) the self-intersecting nodoidlike cylinder (FIG.2(f)). All these shapes have been 
found in numerical solutions of Eq. (5) with $C=0$\cite{zhe}. 
Except the self-intersecting
cases, they all have real correspondence in vitro and in vivo on vesicle shapes\cite{dob,hot,bess}.
The closed shapes ii)-vii) form two separate prolate and oblate ellipsoid branches 
which are bifurcated from a sphere. The oblate ellipsoids 
including CBS have lower energy than the prolate ellipsoids including sphere, and a CBS with 
$c_0 r_v=1.00$ has the lowest energy. The usual 
instability analysis of sphere leads to that the least stable shapes are prolate
and oblate ellipsoid branches\cite{oy89,pet}, but it is the first time to give an explicit
parameterization to show how these two branches come out gradually and analytically. \\
\\
\\
  We are indebted to Professors. Peng Huan-Wu and Zheng Wei-Mou 
for enlightening discussions. This subject is supported by National 
Natural Science Foundation of China.



\begin{figure}
\psfig{file=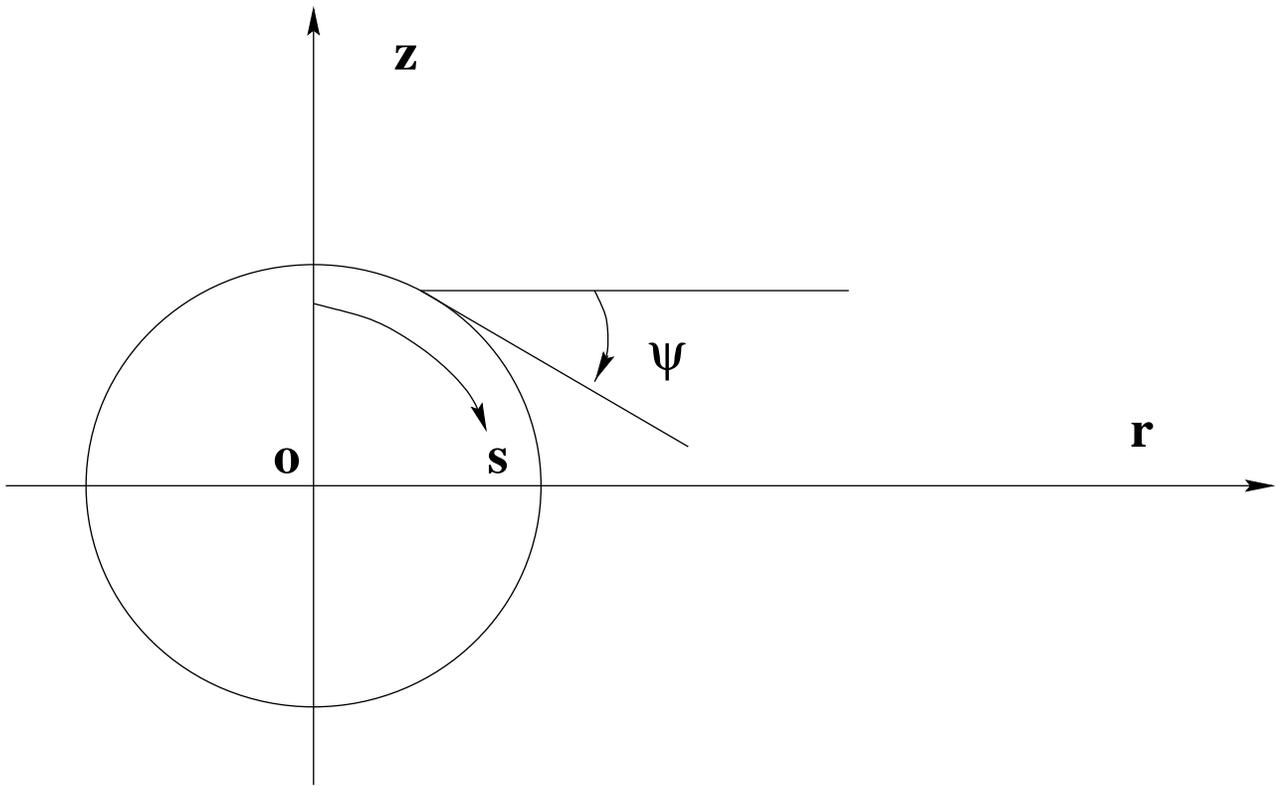,height=105mm,width=170mm,angle=0}
\caption{ Sign convention. Four arrows mean positive directions for rotational axis $z$ and radial axis $r$,
tangent angle {\rm $\psi$} and arclength $s$, respectively.
At the north pole both the arclength $s$ and the tangent angle {\rm $\psi$} take zero values.}
\label{FIG.1}
\end{figure}

\newpage

\begin{figure}
\vspace*{-0.0cm}
\label{FIG 2}
\centerline{
\hbox{\psfig{figure=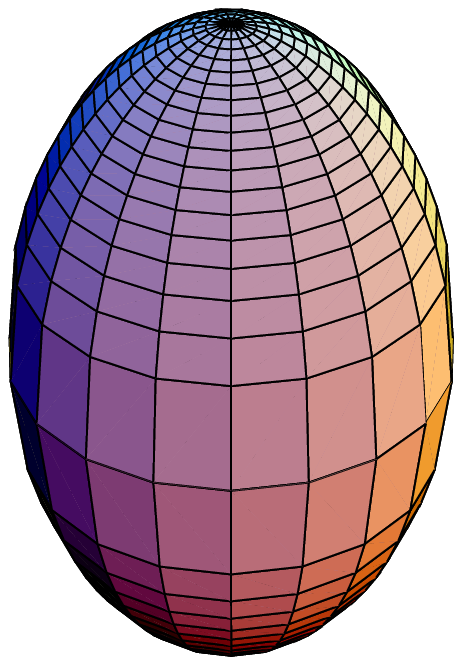,height=4.5cm} \hspace*{2cm} 
      \psfig{figure=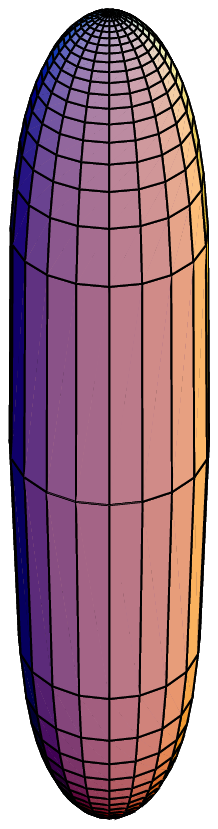,height=8.5cm}}
      }
\centerline{(a)\qquad\qquad\qquad\qquad\qquad(b)}
\centerline{
\hbox{\psfig{figure=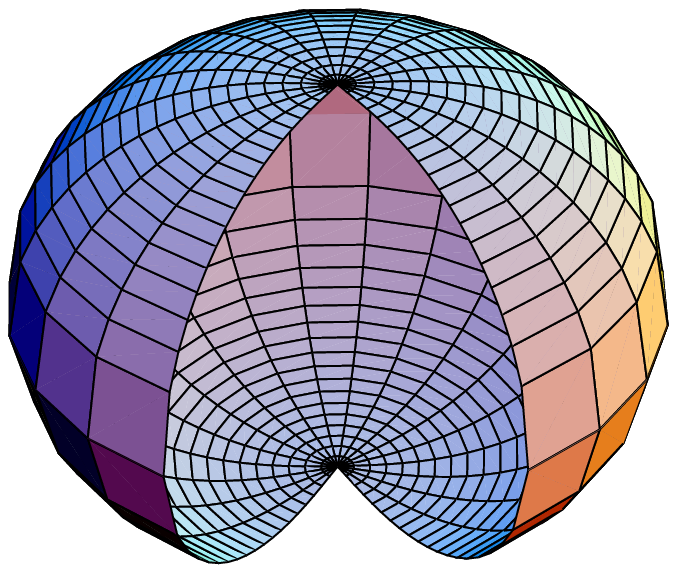,height=4cm}\hfill
      \psfig{figure=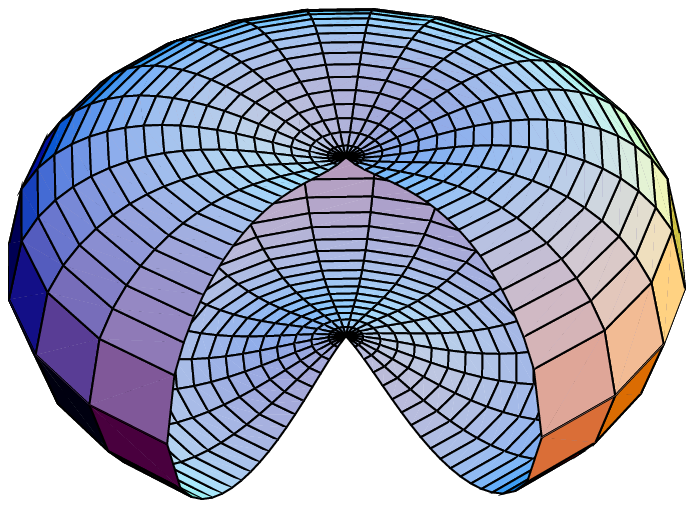,height=4cm}\hfill
      \psfig{figure=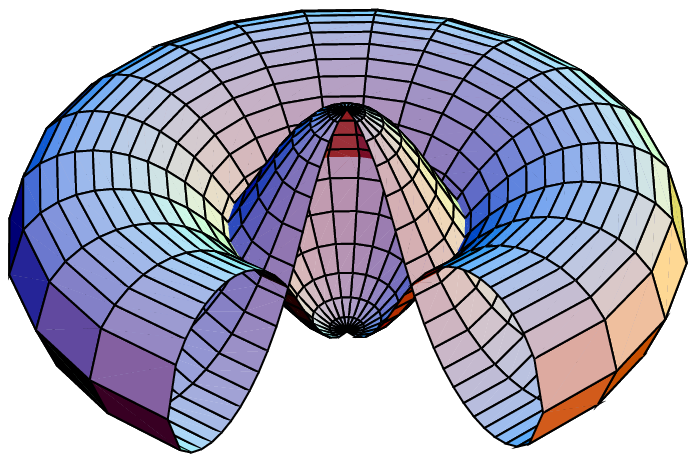,height=4cm}\hfill
      \psfig{figure=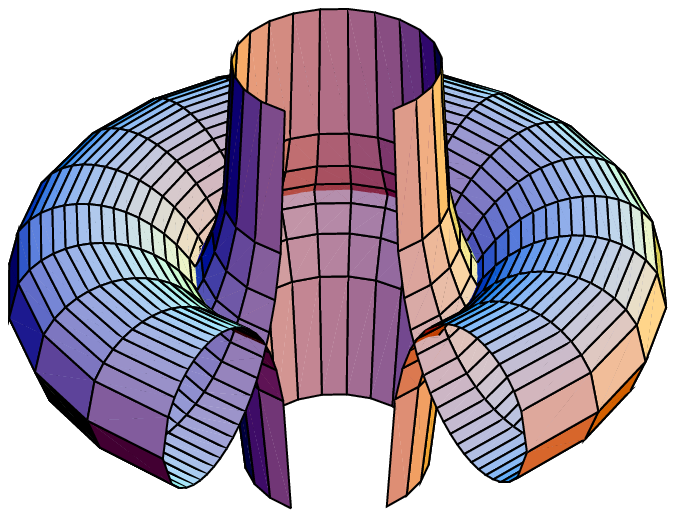,height=4cm}}
      }
\centerline{(c)\qquad\qquad\qquad\qquad\qquad
            (d)\qquad\qquad\qquad\qquad\qquad
            (e)\qquad\qquad\qquad\qquad\qquad(f)}
        \caption{The nontrivial shapes in the CBS solution. The prolate ellipsoid (a), the capped cylinder(b),
        the oblate ellipsoid (c), the CBS (d), the self-intersecting inverted CBS (e) and the self-intersecting nodoidlike cylinder (f).}
\end{figure}

\newpage

\begin{figure}
\psfig{file=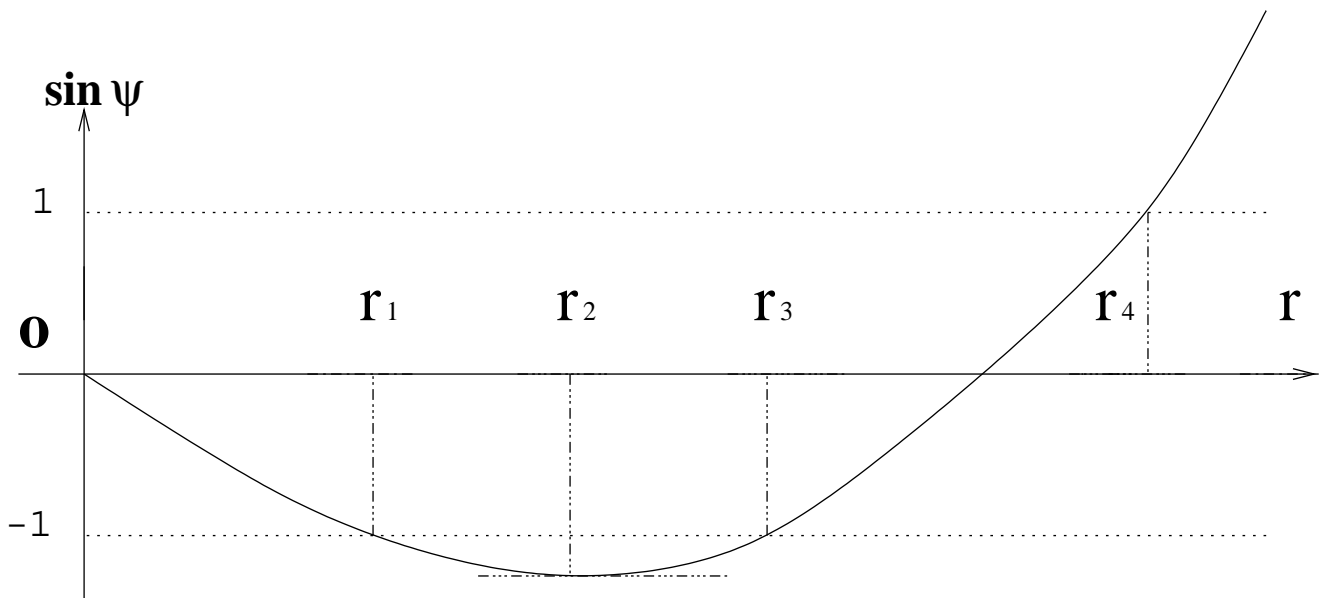,height=80mm,width=175mm,angle=0}
\caption{The relation of $\sin \psi(r)$ vs. $r$ and the intervals where the shapes appear. When 
$0<c_0<c_r$, $-1<\sin\psi(r_2)<0$, shapes appear in $[0,r_4]$. When $c_0\geq c_r$, $\sin\psi(r_2)\leq -1$, shapes appear 
in two distinct intervals $[0,r_1]$ and $[r_3,r_4]$.}
\label{FIG.3}
\end{figure}

\newpage

\begin{figure}
\label{FIG.4}
\vspace*{-0.0cm}
\centerline{\psfig{figure=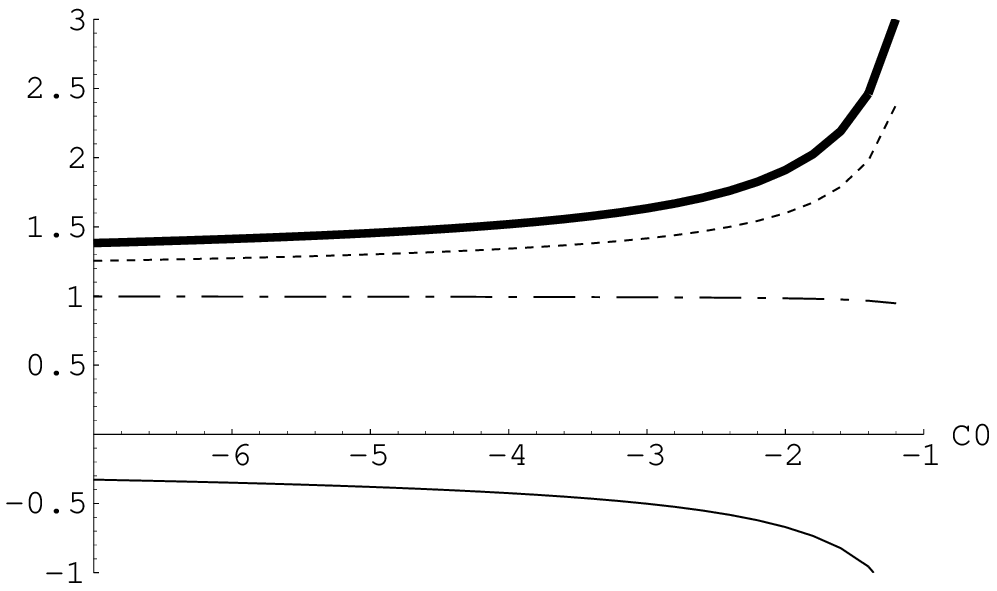,height=4cm}}
\centerline{(a)}
\centerline{\psfig{figure=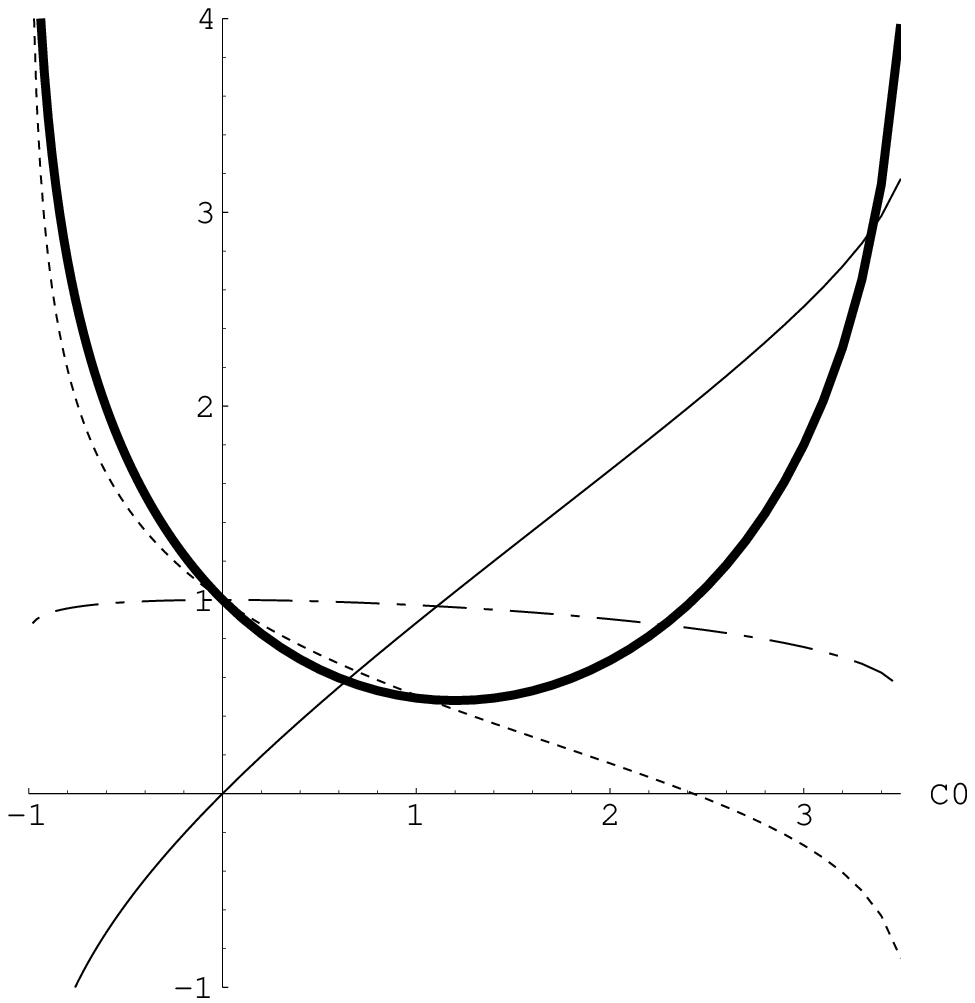,height=6cm}}
\centerline{(b)}
\centerline{\psfig{figure=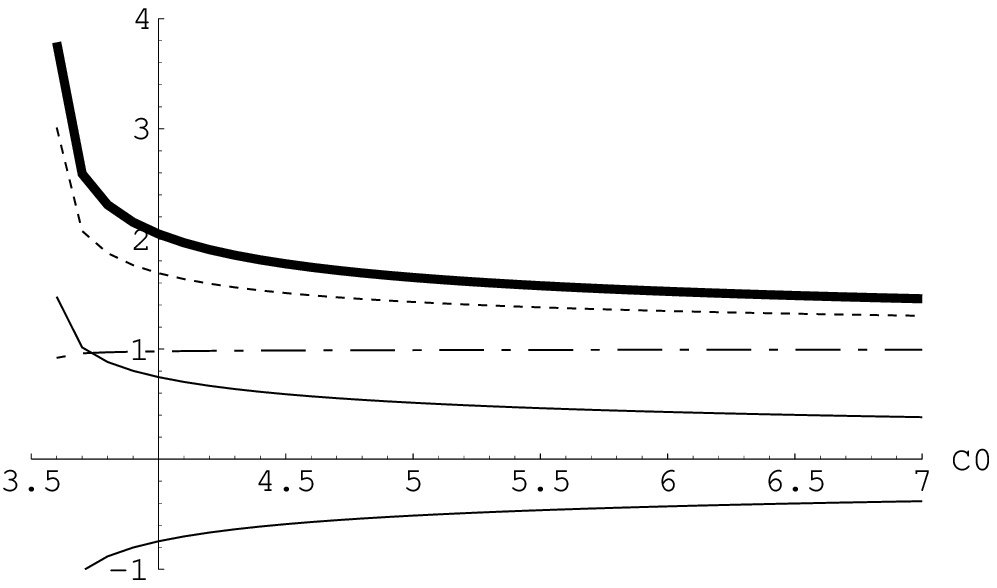,height=4cm}}
\centerline{(c)}
\caption{The energy (thick solid line), the scale invariant $c_0 r_s$ (thin solid line),
two ratios $z(r_0)/r_0$ (dashed line) and $r_v/r_s$ (dot-dashed line),  
in three domains $-7\leq c_0\leq -1.2$ (a), $-0.98\leq c_0\leq 3.5$ (b), and
$3.6\leq c_0\leq 7$ (c), for closed shapes. The bending elastic modulus $k$ used to scale the energy
of a sphere to unit. All quantities in four curves are dimensionless. In order to see clearly that fact that a shape 
with an exclusive value of $c_0\in (-\infty,-1)$ and a shape with an exclusive value of $c_0\in (c_r, \infty)$ 
are identical except that the normals
point inwards and outwards respectively, in last figure (c), we plot both $c_0 r_s$ (above the $c_0$ axis) 
and $-c_0 r_s$ (below the $c_0$ axis ) by the same thin solid lines. We can also see that 
all shapes with $c_0\in (-\infty,-1)$ and $c_0\in (c_r,\infty)$ have their corresponding identical 
shapes with $c_0\in (-1, 0)$. A single domain $c_0\in (-1, c_r)$ thus suffices to give all shapes.}
\end{figure}

\end{document}